\documentclass[preprint1]{aastex}

\newcommand{\half}{{\textstyle{1\over2}}}

\newcommand{\be}{\begin{equation}}
\newcommand{\ee}{\end{equation}}
\newcommand{\bea}{\begin{eqnarray}}
\newcommand{\eea}{\end{eqnarray}}

\def\bi{\bf}

\newfont{\myfont}{cmmib10}
\newcommand{\brho}{\hbox{\myfont \symbol{26} }}

\shorttitle{Faraday rotation}
\shortauthors{Melrose}

\begin{document}

\title{Faraday rotation: effect of magnetic field reversals}

\author{D. B.  Melrose}

\affil{SIfA, School of Physics, University of Sydney, NSW 2006, Australia}

\begin{abstract}
The standard formula for the rotation measure, RM, which determines the position angle, $\psi={\rm RM}\lambda^2$, due to Faraday rotation, includes contributions only from the portions of the ray path where the natural modes of the plasma are circularly polarized. In small regions of the ray path where the projection of the magnetic field on the ray path reverses sign (called QT regions) the modes are nearly linearly polarized. The neglect of QT regions in estimating RM is not well justified at frequencies below a transition frequency where mode coupling changes from strong to weak. By integrating the polarization transfer equation across a QT region in the latter limit, I estimate the additional contribution $\Delta\psi$ needed to correct this omission. In contrast with a result proposed by \cite{BB10}, $\Delta\psi$ is small and probably unobservable. I identify a new source of circular polarization, due to mode coupling in an asymmetric QT region. I also identify a new circular-polarization-dependent correction to the dispersion measure at low frequencies.
\end{abstract}


\section{Introduction}

Faraday rotation is the rotation of the position angle, $\psi$, of linear polarization of radiation passing through a magnetized plasma whose natural modes are circularly polarized. The handedness of the polarization of the natural modes is an odd function of $\cos\theta=\sigma_V|\cos\theta|$, where $\theta$ is the angle between the ray path and the magnetic field. The sense of Faraday rotation reverses when the sign $\sigma_V$ reverses. The net amount of Faraday rotation is proportional to the square of the wavelength, $\lambda$, and is written ${\rm RM}\lambda^2$, defining the rotation measure, RM. The conventional formula for RM, cf.\ (\ref{RM2}) below, has positive and negative contributions from the portions of the ray path with $\sigma_V=\pm1$, respectively. This standard formula neglects the contributions of the small regions around the points where $\sigma_V$ changes sign, and $\theta$ passes through $\pi/2$.  As such a magnetic-field reversal is approached the polarization of the natural modes becomes increasingly elliptical, as $|\cos\theta|$ decreases, becoming linear at $\theta=\pi/2$. Mode coupling needs to be taken into account in such regions, and there are two limiting cases. In the limit of strong mode coupling, the width of the region where the modes are not circularly polarized is negligible, and it can be treated as a sharp boundary.  In the opposite limit of weak mode coupling the components in the two modes remaining in those modes, evolving from circular through elliptical to linear at $\theta=\pi/2$, through elliptical back to circular in the opposite sense. The standard formula for RM effectively assumes strong mode coupling. The strength of mode coupling decreases with decreasing frequency, and below a transition frequency, $\nu_T$, mode coupling becomes weak. The motivation for the present investigation arose in connection with potentially observable effects on RM for $\nu\lesssim\nu_T$. A new result that emerges from the investigation is that partial mode coupling can result in incident linearly polarized radiation emerging with a circularly polarized component.

There is an extensive literature on mode coupling in connection with solar radiophysics, e.g., as summarized is several monographs \citep{B66,Z70,M80,B02}. Although the jargon used is outdated and confusing, I use aspects of it here. Specifically, the approximations in which the natural modes are nearly circularly polarized and nearly linearly polarized are referred to as the QL approximation and the QT approximation, respectively. The region about the point on a magnetic field reversal where the modes satisfy the QT approximation is referred to as a QT region; the center of the QT region is at $\theta=\pi/2$.  In the solar case the radiation of interest is initially circularly polarized. In the strong coupling limit an initially circularly polarized component preserves its polarization on crossing a QT region, corresponding to flipping from one mode to the other. In the weak coupling limit, an incident circularly polarized component follows the polarization of the mode it is in, emerging from the QT region with the opposite handedness. Intermediate mode coupling causes partial conversion of circular into linear polarization, leading to a net depolarization \citep{ZZ64}. 

In the case of Faraday rotation, the initial polarization is linear and corresponds to an equal mixture of two circularly polarized modes. In a recent paper, \cite{BB10} (hereafter BB) discussed the contribution of QT regions to  Faraday rotation in the limit of weak mode coupling, which BB referred to as the  ``super-adiabatic'' (SA) regime. BB argued that a different formula for RM applies in the SA limit, cf. (\ref{RM1}) below, and they suggested that observation of this effect would provide new information on the properties of the interstellar medium (ISM). Part of BB's justification for their new formula (\ref{RM1}) is that a ``reflection'' in the sense $\psi\to-\psi$ occurs at each QT region in the SA limit. This is not correct.  BB's assumption that the position angle of the radiation incident on the QT region can be specified arbitrarily, by $\psi=\psi_{\rm in}$ say, turns out to be incompatible with the constraint that the incident radiation be an equal mixture of the two modes. This constraint determines $\psi$ at the center of the QT region, and the transfer equation determines $\psi$ everywhere else along the ray path (in the SA limit). One is not free to specify $\psi_{\rm in}$ arbitrarily at any point identified as the boundary of the QT region: it is predetermined by the value at the center of the QT region and the details of the model. A more qualitative argument suggests that $\psi\to-\psi$ cannot be correct. Suppose there is a net total of $N$ complete rotations along the ray path when the first QT region is encountered, implying $\psi=\psi_{\rm in}+\pi N$, where for the sake of discussion I assume that $\psi_{\rm in}<\pi$ can be specified. The QT region has no ``knowledge'' of the number $N$, and cannot reverse this through $\psi_{\rm in}+\pi N\to-\psi_{\rm in}-\pi N$, as BB's $\psi\to-\psi$ requires. The actual evolution of $\psi$ through a QT region (in the SA limit) is described in detail below.

My purpose in this paper is to give a critical review of the effect of a QT region on Faraday rotation, and its generalization to the case where the modes are not circularly polarized. A specific motivation is to justify the foregoing comments on BB's treatment. More generally, I consider the possibility that a QT region can induce a net circular polarization, and find that this is indeed possible. Most of my detailed discussion is restricted to the limit of weak mode coupling (SA limit), which is seemingly simple: the radiation initially in one mode remains in that mode and the polarization of that mode at any point is determined by the local parameters. An important point that distinguishes the case of Faraday rotation from other examples of mode coupling is the constraint that the initial radiation is an equal mixture of the two (oppositely circularly polarized) modes. This constraint leads to the surprising (to me) implication that $\psi_{\rm in}$ cannot be arbitrarily specified. I describe the transfer of polarized radiation using a matrix equation for the Stokes parameters, $I,Q,U,V$ \citep{MM91}. This formalism does not involve separating into the two natural modes, and can be used without any reference to modes or mode coupling. The properties of the modes are implicit and can be made explicit by identifying the four eigenvalues of the transfer matrix, and constructing the four eigenfunctions (combinations of $I,Q,U,V$). Two of the eigenvalues are zero, and the other two correspond to the natural wave modes.  A zero eigenvalue implies a conserved quantity, and $I=$ constant is one of these. The other conserved quantity is determined, for Faraday rotation across a QT region, by the constraint that the radiation be an equal mixture of the two modes. To treat Faraday rotation  in the SA limit I derive a differential equation for $\psi=\half\arctan(U/Q)$ subject to this constraint.

Some background results are summarized in Section~\ref{sect:phases}; polarization transfer is discussed in Section~\ref{sect:transfer}; the existence of circular polarization due to mode coupling at a QT region is pointed out in  Section~\ref{sect:circular}; the possibility of observing the effects identified is discussed in Section~\ref{sect:obs}, and the conclusions are summarized in Section~\ref{sect:concl}.

\section{Phases, modes and the Faraday angle}
\label{sect:phases}

In this section I write down results relating to Faraday and mode coupling that are needed in the subsequent discussion.

\subsection{Standard formula for RM}

The standard formula for RM can be written the form, BB's Eqn (15),
\be
{\rm RM}=C\int n_e{\bf B}\cdot d{\bf\ell},
\label{RM2}
\ee
where ${\bf B}\cdot d{\bf\ell}$ is proportional to $\cos\theta$, and where $C$ depends on fundamental constants.  The integral in (\ref{RM2}) is along the ray path through the QL regions. The contribution of the  QT regions is neglected. 

Although the derivation of (\ref{RM2}) is well known, there are several points worth emphasizing. Formula (\ref{RM2}) may be derived by separating initially linearly polarized radiation into two equal circularly polarized  components, corresponding to the two natural wave modes, labeled~o and~x say. The difference in refractive index, $n_o-n_x>0$, causes the two components to get out of phase at a rate $d(\phi_{\rm o}-\phi_{\rm x})/d\ell=\Delta k=(n_{\rm o}-n_{\rm x})\omega/c$ per unit distance, $\ell$, along the ray path. An important point is that the handedness of the~o and~x modes depends on the sign $\sigma_V=\cos\theta/|\cos\theta|$, and that Faraday rotation results from the difference in the refractive indices of the right (R) and left (L) handed natural modes, which is $n_{\rm R}-n_{\rm L}=\sigma_V(n_{\rm o}-n_{\rm x})$. Within a QL region, the position angle rotates at the rate $d\psi/d\ell=\half\sigma_V\Delta k$, which integrates to give (\ref{RM2}). The sign $\sigma_V$ is opposite  in successive QL regions, with the reversal occurring at the center of the QT region that separates the successive QL regions.

\subsection{BB's formula for RM in the SA limit}

The new formula proposed by BB to take account of the effect of the QT regions in the SA limit is, BB's Eqn (14),
\be
{\rm RM}_{\rm SA}=C\int n_e|{\bf B}\cdot d{\bf\ell}|.
\label{RM1}
\ee
As in (\ref{RM2}) the integral is along the ray path through the QL regions. The contributions of the QT regions do not appear explicitly in (\ref{RM1}). 

I give the following physical interpretation of (\ref{RM1}). If the phase difference between the o~and x~modes  is integrated along the ray path through the QL regions, neglecting the QT regions, the accumulated phase difference is ${\rm RM}_{\rm SA}\,\lambda^2/2$, with ${\rm RM}_{\rm SA}$ given by (\ref{RM1}). Suppose that the SA limit applies at every QT region along the ray path. Then  the two components defined by the initial separation into modes remain in these modes everywhere along the ray path, including every QT region. In this case the accumulated phase difference between the o~mode and x~mode components corresponds to the actual accumulated phase difference that develops between the two components defined by the initial separation at the source. It is tempting to think that in the SA limit this accumulated phase difference is relevant to Faraday rotation, as  (\ref{RM1}) appears to imply. However, Faraday rotation is determined by the accumulated phase difference between the R~and L~modes, and not by that between the o~and x~modes. What happens in the QT regions does not affect the fact that the contribution of the QL regions is given by (\ref{RM2}), provided only that the contribution of the QT regions is ignored. BB's argument for (\ref{RM1}) in the SA is that the contribution of each QT region is to cause $\psi\to-\psi$, which effectively converts (\ref{RM2}) into (\ref{RM1}). 

\subsection{Mode coupling}

Traditionally, the strength of mode coupling is described by a coupling coefficient, $Q_x$, such that mode coupling is strong for $Q_x\gg1$ and weak for $Q_x\ll1$. The theory of mode coupling leads to \citep{C60}
\be
Q_x=\left({\nu\over\nu_T}\right)^4,
\qquad
\nu_T=\left({\pi\over2}{\nu_p^2\nu_B^3L\over c}\right)^{1/4},
\label{mc1}
\ee
where $L$ is a characteristic length over which $\theta$ changes, $\nu_p$ is the plasma frequency and $\nu_B$ is the electron cyclotron frequency. The strong dependence on $\nu/\nu_T$ suggests that the transition from strong to weak mode coupling occurs rapidly as a function of decreasing $\nu$.

\section{Transfer of polarized radiation}
\label{sect:transfer}

In this section I present the matrix method for treating the transfer of polarized radiation in the SA limit and apply it to Faraday rotation.

\subsection{Transfer of the Stokes vectors}

The transfer of polarized radiation between the source and the observer can be described by the transfer equation for the Stokes parameters, $I,Q,U,V$. In the present case, $I$ is a constant. The transfer equation for the polarization can be written in the matrix form \citep{MM91}:
\be
{d\over d\ell}\left(
\begin{array}{c}
Q\\U\\V
\end{array}
\right)=
\left(
\begin{array}{rrr}
0\;\;&-\rho_V&\rho_U\\
\rho_V&0\;\;&-\rho_Q\\
-\rho_U&\rho_Q&0\;\;
\end{array}
\right)
\left(
\begin{array}{c}
Q\\U\\V
\end{array}
\right).
\label{gFr}
\ee
The parameters $\rho_Q,\rho_U,\rho_V$ are derived from the (cold plasma) dielectric tensor for the plasma. The transfer equation (\ref{gFr}) is derived for constant $\rho_Q,\rho_U,\rho_V$, and it remains valid provided that any variation in $\rho_Q,\rho_U,\rho_V$ is slow compared to the rate, $\Delta k$, at which the modes get out of phase.

An alternative way of writing (\ref{gFr}) involves interpreting ${\bf S}=(Q,U,V)$ and $\brho=(\rho_Q,\rho_U,\rho_V)$ as vectors in a 3-dimensional vector space. Then (\ref{gFr}) becomes $d{\bf S}/d\ell=\brho\times{\bf S}$. This alternative form implies two conserved quantities, ${\bf S}^2=Q^2+U^2+V^2$ and $\brho\cdot{\bf S}=\rho_QQ+\rho_UU+\rho_VV$. In a slowly varying anisotropic medium, $\brho$ varies slowly in both magnitude and direction. The adiabatic invariants are then ${\bi S}^2/I^2$ and $\brho\cdot{\bf S}/\rho I$, with $\rho=|\brho|=\Delta k$. ($I$ varies as the refractive index varies, but this is insignificant here.) The invariant $\brho\cdot{\bf S}/\rho$ has important implications for Faraday rotation in the SA limit.

No separation into modes is made in deriving (\ref{gFr}), and no separation into modes is needed to use it to describe Faraday rotation. In principle, given a model for the medium, one can integrate (\ref{gFr}) along the ray path from the source to the observer. Some examples of numerical solutions of (\ref{gFr}) for the change in Stokes parameters across a QT region for initially circular polarization were presented by \cite{MRF95}. Analytically, one may integrate the matrix equation (\ref{gFr}) to construct a Mueller matrix, which formally solves the transfer problem  \citep{M92,MJ04}.  Here the specific interest is in the position angle $\psi$, and it is more convenient to use (\ref{gFr}) to derive a differential equation for $\psi$. 

\subsection{Natural wave modes}

The properties of natural wave modes are implicit in the square matrix in (\ref{gFr}), and can be derived in terms of the eigenvalues and eigenfunctions of this matrix. The two non-trivial eigenvalues of the square matrix in (\ref{gFr}) are $\pm i\rho$, $\rho=(\rho_Q^2+\rho_U^2+\rho_V^2)^{1/2}$ and the eigenvectors  correspond to the two natural modes. The polarization of the natural modes may be described by the axial ratio, $T=\sigma_V|T|$, of the o~mode, and the orientation $\psi_B$ of its major axis. For the x~mode the axial ratio is $-1/T$ at $\psi_B+\pi$. One has \citep{MM91}
\be
\rho_Q=-\Delta k\,\cos2\chi_B\,\cos2\psi_B,
\qquad
\rho_U=-\Delta k\,\cos2\chi_B\,\sin2\psi_B,
\qquad
\rho_V=-\Delta k\,\sin2\chi_B,
\label{rhoQUV}
\ee
with
\be
\Delta k=(\rho_Q^2+\rho_U^2+\rho_V^2)^{1/2},
\qquad
\cos2\chi_B=
{T^2-1\over T^2+1},
\qquad
\sin2\chi_B={2T\over T^2+1}.
\label{rhoQUV1}
\ee
Approximations to $\Delta k$ in terms of the plasma parameters are
\be
\Delta k={\pi\nu_p^2\nu_B\over\nu^2c}
\left\{
\begin{array}{ll}
|\cos\theta|,&\quad |\cos\theta|\gtrsim\nu_B/2\nu,\\
\nu_B/2\nu,&\quad |\cos\theta|\lesssim\nu_B/2\nu.
\end{array}
\right.
\label{Deltak}
\ee
The case $|\cos\theta|\lesssim\nu_B/2\nu$ applies within a QT region, where $\Delta k$ is independent of $\theta$.

\subsection{A conserved quantity}

As already noted, $\rho_QQ+\rho_UU+\rho_VV$ is conserved by (\ref{gFr}). In the application to Faraday rotation, the initial condition corresponds to linearly polarized radiation, implying $V=0$, and circularly polarized modes, implying $\rho_Q=\rho_U=0$, and hence $\rho_QQ+\rho_UU+\rho_VV=0$. It is convenient to write $(Q,U,V)=I(\cos2\chi\,\cos2\psi,\cos2\chi\,\sin2\psi,\sin2\chi)$. The conserved quantity for Faraday rotation becomes
\be
\cos2\chi_B\,\cos2\chi\,\cos2(\psi-\psi_B)+\sin2\chi_B\,\sin2\chi=0.
\label{con}
\ee
Within a QT region, the modes are elliptically polarized, and the radiation becomes elliptically polarized, implying that both terms in (\ref{con}) are nonzero, so that (\ref{con}) constrains the angle, $\psi-\psi_B$, between the orientation of the polarization ellipse of the radiation and that of the o~mode. At the center of the QT region, where the modes are linearly polarized, $\sin2\chi_B=0$ requires $\cos2(\psi-\psi_B)=0$, implying $\psi-\psi_B=\pm\pi/4$. 

The conserved quantity (\ref{con}) has the surprising implication that one is not free to specify the position angle, $\psi_{\rm in}$, of radiation incident on a QT region. One may interpret this as follows. The angle $\psi_B$ is indeterminate in the limit of circular polarization, and as a QT region is entered and the modes become significantly elliptical, the angle $\psi_B$ becomes relevant in determining the orientation of the polarization ellipses of the two natural modes. The requirement that the radiation be an equal mixture of the two modes determines the angle $\psi$ in terms of the $\psi_B$ (and  $\chi_B$). Hence, $\psi$ is predetermined by the assumption that the radiation is an equal mixture of the two modes, and there is no freedom to specify $\psi_{\rm in}$.

\subsection{Faraday angle within a QT region}

One may use (\ref{gFr}) to derive coupled equations that describe the changing ellipticity, described by $\chi$, and position angle, $\psi$, within a QT region. The ellipticity changes according to
\be
{d\chi\over d\ell}=\half\rho\cos2\chi_B\,\sin2(\psi-\psi_B).
\label{chiell}
\ee
The position angle changes according to
\be
{d\psi\over d\ell}=\half\rho[\sin2\chi_B
-\cos2\chi_B\,\tan2\chi\,\cos2(\psi-\psi_B)]=\half\rho\,{\sin2\chi_B\over\cos^22\chi},
\label{psiell}
\ee
where the second form follows by using (\ref{con}).

\subsection{Model for a QT region}

A model for a QT region needs to specify how the plasma parameters vary with distance, $\ell$, along the ray path. Relevant parameters are the angles that describe the orientation of the magnetic field with respect to the ray path. One has $\cos\theta=0$ at the center of the QT region, at $\ell=\ell_1$ say, corresponding to $\sin2\chi_B=0$, and the model needs to specify how $\cos\theta$ and hence $\sin2\chi_B$ vary on either side of this zero. In a symmetric model $\sin2\chi_B$ is an odd function of $\ell-\ell_1$.  A simple example of a model is $\chi_B={\dot\chi}_B(\ell-\ell_1)$ with ${\dot\chi}_B$ constant near $\ell=\ell_1$. An untwisted magnetic field corresponds to $\psi_B=$ constant. For a twisted magnetic field the model must also specify how $\psi_B$ varies as a function of $\ell$.  An analogous simple model is $\psi_B={\dot\psi}_B(\ell-\ell_1)$ with ${\dot\psi}_B$ constant near $\ell=\ell_1$.  Mode coupling is strong when either ${\dot\chi}_B$ or ${\dot\psi}_B$ is large compared with $\Delta k$, and mode coupling is weak when both are small compared with $\Delta k$.

\begin{figure} [t]
\centerline{
\includegraphics[width=10cm]{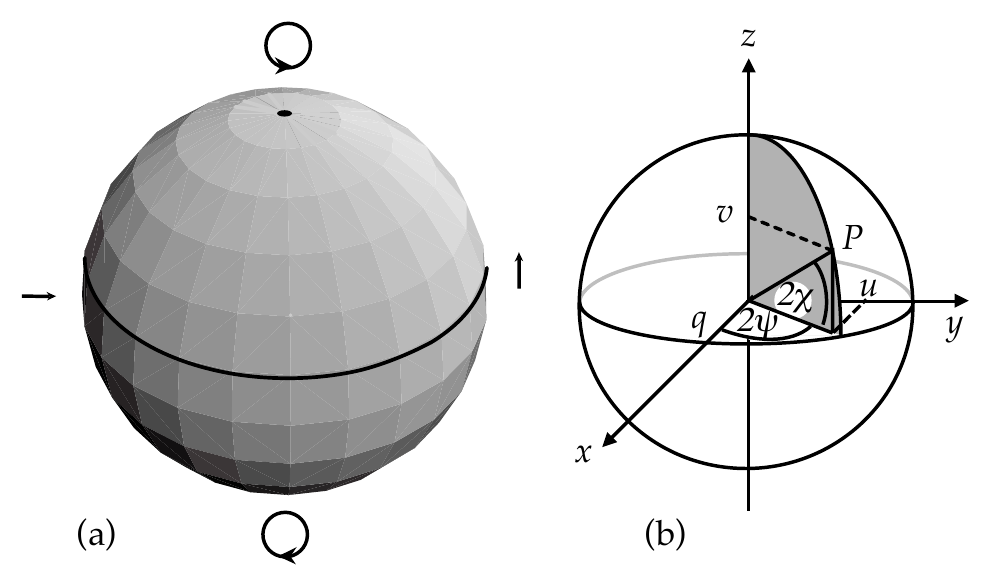}
}
\caption{An arbitrary polarization, represented in terms of Stokes parameters by $q=Q/I$, $u=U/I$, $v=V/I$ is represent on the Poincar\'e sphere by colatitute $2\chi$ and longitude $2\psi$, with $\psi$ describing the plane of linear polarization. [After \cite{MM91}]}
\label{fig:Poincare}
\end{figure}

\subsection{Motion on the Poincar\'e sphere}

Interpretation of the equations (\ref{chiell}) and (\ref{psiell}) that determine the evolution of $\chi$ and $\psi$ is facilitated by a pictorial description of the evolution of the polarization on the Poincar\'e sphere, as illustrated in Fig.~\ref{fig:Poincare}. An arbitrary polarization is represented by a point on the sphere, with the two circular polarizations corresponding to the two poles, and linear polarization to the equator. The angle $2\psi$ corresponds to longitude and the angle $2\chi$ to colatitude. The orthogonal polarizations of the two modes correspond to antipodal points, and define a ``mode axis'' through the sphere, with end points at $2\chi_B,2\psi_B$ and $\pi-2\chi_B,2\psi_B+\pi$. The evolution described by (\ref{gFr}) or (\ref{chiell}) and (\ref{psiell}) corresponds to the polarization point rotating relative to the mode axis. Outside QT regions, the modes are circular, and the mode axis is either north-south or south-north. The direction of the mode axis reverses each time a QT region is crossed. In the limit of strong mode coupling the polarization point is approximately stationary during the reversal of the mode axis. In the SA limit the reversal of the mode axis is slow, with the polarization point rotating rapidly about it many times as it moves from, say, north-south to south-north.

The interpretation of the conserved quantity (\ref{con}) is that the polarization point rotates in a great circle about the mode axis. This constraint applies specifically only to Faraday rotation in the SA limit, and expresses the requirement that the radiation remain an equal mixture of the two modes.

In discussing Faraday rotation, the quantity of interest is $\psi$, which varies periodically at a rate $\Delta k$. In a QL region, the great circle is around the equator, and the polarization point moves at constant velocity, $\Delta k$, around it. As the QT region is entered, the mode axis moves away from south-north, and the great circle tilts. According to (\ref{psiell}), the variation in $\psi$ continues with the same period, but with a non-uniform velocity, being slower near the points where the great circle crosses the equator and faster where the great circle is furthest from the equator. This non-uniform $d\psi/d\ell$ reflects uniform circular motion around the great circle projected onto the equatorial plane, where the projection is an ellipse with angle $2\psi$. As the mode axis approaches the equator, the projection of the great circle onto the equatorial plane becomes highly eccentric.

At the center of the QT region the mode axis is in the equatorial plane, and the great circle passes through both poles. According to (\ref{psiell}), the rate of change of $\psi$ becomes formally infinite for $\sin2\chi=0$, corresponding to the poles. This has a simple interpretation. The motion around the great circle is at constant longitude, which changes abruptly by $\pm\pi$ as a pole is crossed: specifically, $2\psi$ is at one value as the point moves from the south pole to the north pole in the western hemisphere, and changes abruptly by $-\pi$ as it crosses the north pole and enters the eastern hemisphere, where it remains constant until the point reaches the south pole. With the projection of the great circle reducing to a line, the point is stationary at one end point for half the period, moving at infinite speed to the other end point as a pole is crossed.

On crossing the center of the QT region to the other side, the sense of the motion around the ellipse formed by projecting the great circle onto the equatorial plane reverses. If the QT region is symmetric about its center, the increase in $\psi$ on one side of the center is balanced exactly by the decrease in $\psi$ on the other side. Thus, the net effect on $\psi$ of a symmetric (untwisted) QT region is nil. The formula (\ref{RM2}) for Faraday rotation then applies without modification in the SA limit.  A net change $\Delta\psi$ is possible for a QT region only if it is asymmetric in some sense. For an untwisted field, this requires that $\cos\theta$ not be strictly an odd function of $\ell-\ell_1$. The actual value of $\Delta\psi$ then depends on the specific form of this asymmetry.

If the magnetic field is twisted, there is a contribution to $\Delta\psi$, in the SA limit, due to the twist. This is due to the radiation remaining in the natural modes as the orientation of their polarization ellipses follows the changes in longitude, $2\psi_B, 2\psi_B+\pi$, on the Poincar\'e sphere. The contribution to $\Delta\psi$ can be estimated as ${\dot\psi}_B\Delta\ell$, where $\Delta\ell$ is the length of the ray path corresponding to the QT region. Although a detailed calculation based on a specific model is needed to determine $\Delta\ell$, an approximate estimate is the distance between the points where $\tan2\psi_B$ is equal to $\pm1$.

In summary, the periodic variation of $\psi$ continues through the QT region, and reverses sense at the center of the QT region, where the requirement that there be an equal mixture of two modes constrains the great circle to pass through both poles, with $\psi$ having one of two possible values, corresponding to the longitude of the great circle in the eastern and western hemispheres. In a symmetric QT region, the net effect on $\psi$ is zero, implying that no change in (\ref{RM2}) occurs in passing from the limit of strong to the limit of weak mode coupling. Small changes, $\Delta\psi$, can result from an asymmetry in the QT region, including a twist in the magnetic field.

\subsection{Polarization ``reflection'' at a QT region}

The foregoing discussion implies that BB's suggested ``reflection'' $\psi\to-\psi$ at a QT region in the SA limit does not occur. The model used by BB (their Appendix~B) assumes radiation incident on the QT region in an equal mixture of two modes, which I write as $e^{i\Delta\phi/2}{\bf e}_{\rm o}+e^{-i\Delta\phi/2}{\bf e}_{\rm x}$, with $\Delta\phi$ the phase difference between the two modes. Suppose $\rm o=R$, $\rm x=L$ on the incident side, and that the components remain in their initial modes. The handedness of each mode reverses across the QT region, implying $\rm o=L$, $\rm x=R$ for the emerging radiation. With $2\psi=\Delta\phi$ on the incident side, BB interpreted the interchange of the coefficients of the R~and L~polarization vectors as implying the putative polarization reflection, $\psi\to-\psi$. However, my analysis implies that  the assumption of an equal mixture of the two modes requires $\psi-\psi_B=\pm\pi/4$ at the center of the QT region, effectively constraining it (at the center of the QT region) to what BB call ``the $x=y$-axis'' about which their putative reflection occurs. There is no ``reflection'': $\psi$ changes systematically in one sense until the center of the QT region is crossed, and then it changes systematically in the opposite sense.

\section{Circular polarization induced by a QT region}
\label{sect:circular}

In this section I point out that a QT region can lead to a partial conversion of linear into circular polarization. The inverse is well known: circularly polarized radiation incident on a QT region in the intermediate range where mode coupling is neither strong nor weak leads to emerging radiation that is partially linearly polarized \citep{ZZ64}. 

\subsection{Mode coupling across a symmetric QT region}

Mode coupling is determined by a coupling coefficient, $Q_x$, that is effectively the ratio of two rates: that at which the shape or orientation of the polarization ellipse of the natural modes changes, and the rate at which the modes get out of phase. Mode coupling is strong when this ratio is large and weak when it is small. In QL regions, where the polarization of the natural modes does not change, mode coupling is weak. As a QT region is approached, the polarization becomes significantly elliptical and $\Delta k$ decreases: $Q_x$ increases due to both effects. In the case of weak coupling one has $Q_x\ll1$ throughout a QT region. Consider intermediate mode coupling, with $Q_x\gg1$ at the center of the QT region, and $Q_x>1$ for some distance on either side of the center. In this case, weak mode coupling applies to radiation approaching the QT, with $Q_x$ increasing from $\ll1$, and passing through unity, say at $\ell=\ell_-$. Mode coupling is then strong across a region, $\ell_-\le\ell\le\ell_+$ say, that includes the center, $\ell=\ell_1$, of the QT region, becoming weak when $Q_x$ passes through unity at $\ell=\ell_+$.

Consider an idealized model in which the components in the two modes follow the changing polarization of the modes up to $\ell=\ell_-$, at which point the polarization of the radiation is frozen-in, remaining unchanged across the center of the QT region and ceasing to be frozen-in at $\ell=\ell_+$, after which the components in the two modes again follow the changing polarization of the modes. In this model, the condition (\ref{con}) is satisfied for $\ell<\ell_-$. Let the polarization at $\ell=\ell_-$ be described by $\chi=\chi_-$ and $\psi=\psi_-$. The radiation is an equal mixture of the two modes so that (\ref{con}) is satisfied with $\chi=\chi_-$, $\psi=\psi_-$ and $\chi_B=\chi_{B-}$, $\psi_B=\psi_{B-}$. The model implies that $\chi=\chi_-$ and $\psi=\psi_-$ remain constant for $\ell_-<\ell<\ell_+$. 

In terms of the vector notation in which (\ref{gFr}) becomes $d{\bf S}/d\ell=\brho\times{\bf S}$, the invariant $\brho\cdot{\bf S}/\rho$ applies for $\ell<\ell_-$ and for $\ell>\ell_+$. The constant quantity over the range $\ell_-<\ell<\ell_+$ is ${\bf S}={\bf S}_-$, with $\brho$ varying and $\brho=\brho_\pm$ different at $\ell=\ell_\pm$. 

For a symmetric QT region, at $\ell=\ell_+$ one has $\chi_{B+}=-\chi_{B-}$ and $\psi_{B+}=\psi_{B-}+\pi$, corresponding to $\brho_+=-\brho_-$. The condition (\ref{con}) is then satisfied at $\ell=\ell_+$. In the vector notation, one has $\brho_+\cdot{\bf S}=-\brho_-\cdot{\bf S}_-=0$ at $\ell=\ell_+$. In a symmetric model, the decrease in $\psi$ for $\ell>\ell_1$ is a mirror image of the increase for $\ell<\ell_1$. The changes across a symmetric QT region cancel exactly. This implies that the standard formula (\ref{RM2}) for RM applies without modification irrespective of the strength of the mode coupling, provided that the QT region is symmetric.

\subsection{Effect of an asymmetry}

Suppose that the QT region is asymmetric in the sense that at $\ell=\ell_+$ one has $\chi_{B+}\ne-\chi_{B-}$ and/or $\psi_{B+}\ne\psi_{B-}+\pi$, that is, $\brho_+\ne-\brho_-$ in the vector notation. This is obviously the case if there is a twist in the magnetic field, when $\psi_{B+}$ differs from $\psi_{B-}+\pi$ by the angle through which the magnetic field twists between $\ell=\ell_-$ and $\ell=\ell_+$. There can also be an asymmetry causing  $\chi_{B+}$ to differ from $-\chi_{B-}$ due to $\cos\theta$ not being a strictly odd function of $\ell-\ell_1$. 

In the absence of symmetry, the condition (\ref{con}) is not satisfied at $\ell=\ell_+$. To be specific, suppose one has
\be
\cos2\chi_{B+}\,\cos2\chi_-\,\cos2(\psi-\psi_{B+})+\sin2\chi_{B+}\,\sin2\chi_-=\delta_+
\label{con+}
\ee
at $\ell=\ell_+$, where $\chi=\chi_-$, $\psi=\psi_-$ applies to the frozen-in polarization which satisfies (\ref{con}) with $\chi_B=\chi_{B-}$, $\psi_B=\psi_{B-}$. As $\psi_B$, $\chi_B$ change with $\ell>\ell_+$, $\chi$ and $\psi$ vary such that (\ref{con+}) continues to be satisfied with fixed $\delta_+$. In terms of the vector notation, for $\ell>\ell_+$ one has $\brho\cdot{\bf S}/\rho$ constant at its value at $\ell=\ell_+$.

The implication of (\ref{con+}) is that the radiation is an unequal mixture of the two modes for $\ell\ge\ell_+$. For $\ell\gg\ell_+$, in the subsequent QL region, this implies that the radiation is an unequal mixture of two opposite circular polarizations, implying a degree of circular polarization $\delta_+$. The handedness of the resulting polarization depends on the sign $\sigma_V$ and the nature of the asymmetry, e.g., the sense of twist of the magnetic field. One expects it to be random:  for a large number of QT regions the signs $\delta_+>0$ and $\delta_+<0$ would occur with equal probability.

A detailed model is required to estimate how the magnitude of $\delta_+$ varies with frequency. The following arguments suggest that $\delta_+$ is a maximum near the transition frequency, $\nu_T$. Mode coupling becomes stronger with increasing $\nu>\nu_T$, implying that for $\nu\gg\nu_T$ the freezing of the polarization begins in the QT region when the modes are only slightly non-circular. This corresponds to small $\cos\chi_{B-}$, $\sin\chi_-$ and hence small $\delta_+$, with $\delta_+$  decreasing as the mode coupling becomes stronger. At low frequencies, $\nu\ll\nu_T$, there is negligible mode coupling, and negligible freezing-in implies negligible induced circular polarization. The optimum conditions for induced circular polarization occur for $\nu\approx\nu_T$, where $\cos\chi_{B-}$ and $\sin\chi_-$ are not small, and the distance, $\Delta\ell=\ell_+-\ell_-$, is large enough for a twist in the magnetic field to have a significant effect on the orientation of the polarization ellipses of the natural modes.

\section{Observable consequences}
\label{sect:obs}

I comment on the possibility of observing three effects: the change $\Delta\psi$, circular polarization due to mode coupling, and the quantity defined by (\ref{RM1}).

\subsection{Effect of $\Delta\psi\ne0$}

For a change, due to mode coupling changing from  strong to weak, to be observable, the transition frequency, $\nu_T$, must be in the range of observation. BB estimated $\nu_T$ for the ISM, and argued that under favorable circumstances it may be in the range of observational interest, $100\,$MHz--$1\,$GHz. Let me assume that this is the case, and consider what would be observed.

If BB's  suggestion were correct that RM is given by (\ref{RM1}) rather than (\ref{RM2}) in the SA regime, the change in RM with decreasing frequency would be potentially very large, with (\ref{RM2}) applying at high frequency and (\ref{RM1}) at low frequency. However, (\ref{RM1}) is not relevant to Faraday rotation, and any actual change in RM from high to low frequency is much smaller than  (\ref{RM1}) would suggest. A change $\Delta\psi$ due to a twist in the field can be written as $\Delta\psi={\dot\psi}_B\Delta\ell$, where $\Delta\ell\propto\lambda$ is the thickness of the QT region. This effect can be observed by determining RM at short wavelengths, and using it to identify the position angle $\psi_0$ at the source (by taking the limit $\lambda\to0$), and then repeating the procedure at longer and longer wavelengths. The prediction is that one would find that the inferred $\psi_0$ has a contribution $\propto{\dot\psi}_B\lambda$ that depends on the wavelength band used to determine it. Observation of such a change would enable one to determine the transition frequency, $\nu_T$, in (\ref{mc1}). 

A complication is that one expects there to be many QT regions along the line of sight. Statistically, half of the changes would have $\Delta\psi>0$ and half have $\Delta\psi<0$. Hence, if the number of QT regions is large, one expects their cumulative effect to sum to zero, leaving the standard result (\ref{RM2}) essentially unchanged for $\nu\ll\nu_T$.

\subsection{Circular polarization due to mode coupling}

Circular polarization induced by a QT region on incident linearly polarized radiation is a new effect that has not been recognized previously. It has some similarity to scintillation-induced circular polarization \citep{MM00}, but is not the same effect. There is a partial analogy with linear polarization induced by a QT region on incident circular polarization \citep{ZZ64}. A notable feature of the analogy is that both effects are maximized when mode coupling is intermediate between the strong and weak limits. This maximum should be near the transition frequency, $\nu\approx\nu_T$.  Observationally, the effect should be confined to a bandwidth of order $\nu_T$ about $\nu_T$ for each QT region.

For a QT region to induce circular polarization it must be asymmetric, and a plausible cause for the asymmetry is a twist in the magnetic field. One would expect this effect to be observable when there is only one QT region along the ray path. In practice one expects many QT regions, and for a statistically large number of QT regions, the degree of circular polarization from each is random, in the sense that there is no systematic sign, and hence would tend to  average to zero.

\subsection{Comparison with solar application}

There is a long-standing problem relating to bipolar regions in the solar corona, where x~mode radiation from opposite footpoints must be oppositely circularly polarized on emission. The radiation from the more distant footpoint must pass through an extra QT region compared to radiation from the nearer footpoint. A prediction is that for $\nu<\nu_T$, the two footpoints should have the same handedness.  Estimates of $\nu_T$ for propagation through the solar corona suggest sufficiently high values that one should observe this change above about $100\,$MHz. However, the effect is not observed  \citep{WTK92}, suggesting that the value of $\nu_T$ is overestimated. 

In my opinion, the likely resolution of this long-standing problem is that the magnetic field has a small-scale random component, and that what is identified as a single QT region where the mean field passes through $\theta=\pi/2$ consists of a large number of mini-QT regions where the local value of $\theta$ passes through $\pi/2$. The strength of mode coupling depends on $L$ in (\ref{mc1}), and this  is very much smaller for the mini-QT regions than would be estimated for the mean field. A random small-scale component of the magnetic field in the ISM would have an analogous effect, reducing the likely value of $\nu_T$ to much less than the estimated range $100\,$MHz--$1\,$GHz of observational interest. 

Suppose the magnetic field has a systematic twist, as well as a random component. The additional rotation, $\Delta\psi$, and the induced circular polarization, $\delta_+$, each have opposite signs at consecutive mini-QT regions. The number of mini-QT regions is necessarily odd, and the net effect should be of the same order as for a single mini-QT region. This suggests that inclusion of a random component makes it more difficult to observe $\Delta\psi$, which requires $\nu\ll\nu_T$, but has less effect on observing $\delta_+$, which requires $\nu\approx\nu_T$.

\subsection{Polarization-dependent DM}

Formula (\ref{RM1})  is not relevant to Faraday rotation, but the quantity (\ref{RM1}) can be measured in principle through a relative time delay between the o~mode and x~mode components. If the SA limit applies at every QT region between the source and the observer, the components in the two modes propagate independently, and (\ref{RM1}) determines the accumulated phase difference between them. The difference in group velocity between the o~and x~mode components is proportional to $n_{\rm o}-n_{\rm x}$, and this leads to a time delay in the arrival of the oppositely circularly polarized component of a pulse. This relative delay between oppositely polarized components increases $\propto\lambda^3$, with the coefficient $\propto {\rm RM}_{\rm SA}$. Observationally, this implies a correction of opposite sign to the DM for oppositely circularly polarized components in a pulse. Although observing a relative time delay between oppositely circularly polarized components is possible in principle, it is essential to have a well-defined pulse to detect. A potential complication is dispersive smearing, which tends to smear out a pulse, with this smearing increasing as the frequency decreases.

\section{Conclusions}
\label{sect:concl}

This investigation of the effect of mode coupling at a QT region on Faraday rotation was motivated by a recent discussion by BB, who argued that at low frequencies, where mode coupling is weak, the standard formula (\ref{RM2}) for Faraday rotation is replaced by (\ref{RM1}). This is based on a misinterpretation of what happens at a QT region;  (\ref{RM1}) is not relevant to Faraday rotation. The analysis given here implies that the standard formula (\ref{RM2}) applies irrespective of the strength of mode coupling at QT regions, provided that each QT region is symmetric. An asymmetry can induce a small correction, $\Delta\psi$, that needs to be added to RM$\,\lambda^2$ in the SA limit. The sign of this correction is essentially random, and the net correction over many QT regions should average to zero.

There are three other new results to emerge from the present investigation, one that is theoretical, and two of possible observational relevance.

The theoretical result relates to the limit of weak mode coupling (BB's SA limit), when the polarization transfer equation (\ref{gFr}) implies an invariant $(\rho_QQ+\rho_UU+\rho_VV)/\Delta k\,I$. The new result is that this invariant is equal to zero for Faraday rotation; this particular value corresponds to an equal mixture of the two modes. One implication is that the polarization point on the Poincar\'e sphere moves around a great circle, whose orientation changes as the polarization of the natural modes changes. A particular implication follows by considering this motion at the center of the QT region where the natural modes are linear: the great circle then passes through both poles, corresponding to cyclic conversion of linear to circular polarization as in a quarter-wave plate. The implication is that the position angle, $\psi$, at the center of the QT region is predetermined by this invariant and the orientation of the magnetic field at that point. This invalidates the simple model used by BB to justify (\ref{RM1}) for Faraday rotation because this model pre-supposes that the initial orientation of the plane of polarization is arbitrary. More specifically, the reflection, $\psi\to-\psi$, about the ``$x=y$-axis'' that BB argued occurs at a QT region is a misinterpretation: the ``$x=y$-axis'' is the position angle of the great circle at the center of the QT region.

The development of circular polarization at a QT region can be understood in terms of this invariant. At intermediate frequencies, $\nu\approx\nu_T$, mode coupling is weak on the edges of a QT region, and strong in a region about its center.  The quantity $(\rho_QQ+\rho_UU+\rho_VV)/\Delta k\,I$ is invariant only when mode coupling is weak. When mode coupling is strong, across a region about the center of the QT region, the polarization of the radiation is frozen-in. Any asymmetry across this region, such as a twist in the magnetic field,  implies that the values of $(\rho_QQ+\rho_UU+\rho_VV)/\Delta k\,I$ at the two points where the transition occurs between weak and strong mode coupling are not equal.  An initial value of zero implies a non-zero final value, and hence an unequal final mixture of the two modes. On emerging from the QT region this becomes an unequal mixture of two circular components, giving a net degree of circular polarization, $\delta_+$. This induced circular polarization should be a maximum near the transition frequency, $\nu_T$. 

Although observation of the effects described here by $\Delta\psi$ and $\delta_+$ encounter difficulties, these may be less serious for $\delta_+$, which requires $\nu\approx\nu_T$, than for $\Delta\psi$, which requires $\nu\ll\nu_T$. The signs of both $\Delta\psi$ and $\delta_+$ depend on details that suggest that they should average to zero over a larger number of QT regions. Realistically, the effects could be observed only if at most a few QT regions contribute.

The quantity defined by the alternative formula (\ref{RM1}) proposed by BB for RM, although not relevant to Faraday rotation, is observable in principle as a relative time delay between the o~mode and x~mode components in the SA limit. Observationally, this effect implies a correction of opposite sign to the DM for oppositely circularly polarized components in a pulse.  Any measurement of this effect would give intrinsically new information of the properties of the ISM. Because this effect favors low frequencies, dispersive smearing is likely to be a problem in any attempt to measure it for pulsars. 

\acknowledgements

I thank Alex Judge for helpful discussions.


\begin{thebibliography}{22}

\bibitem[\protect\citeauthoryear{Benz}{2002}]{B02}
Benz, A. O. 2002, {\it Plasma Astrophysics}, Kluwer Academic Publications, Dordrecht, p.~269

\bibitem[\protect\citeauthoryear{Budden}{1966}]{B66}
Budden, K. G. 1966, {\it Radio waves in the ionosphere}, Cambridge University Press, chapters 18 \& 19

\bibitem[\protect\citeauthoryear{Broderick \& Blandford}{2010}]{BB10}
Broderick, A. E., \& Blandford, R. D. 2010 ApJ 718, 1085

\bibitem[\protect\citeauthoryear{Cohen}{1960}]{C60}
Cohen, M. H. 1960 ApJ 131, 664

\bibitem[\protect\citeauthoryear{Macquart  \& Melrose}{2000}]{MM00}
Macquart, J.-P., \& Melrose, D. B.
{2000}, Phys. Rev. E  62, 4177

\bibitem[\protect\citeauthoryear{Melrose}{1980}]{M80}
Melrose, D. B.  1980, {\it Plasma Astrophysics, Vol. 2},
Gordon \& Breach. New York, chapter 12


\bibitem[\protect\citeauthoryear{Melrose}{1992}]{M92}
Melrose, D. B. 1992, J. Plasma Phys. 50, 267

\bibitem[\protect\citeauthoryear{Melrose \& Judge}{2004}]{MJ04}
Melrose, D. B., \& Judge, A. C. 2004 Phys. Rev. E  70,  056408

\bibitem[\protect\citeauthoryear{Melrose \& McPhedran}{1991}]{MM91}
Melrose, D. B., \& McPhedran, R. C.
{1991}
{\it Electromagnetic Processes in Dispersive Media},
Cambridge University Press, chapter 14

\bibitem[\protect\citeauthoryear{Melrose, Robinson \& Feletto}{1995}]{MRF95}
Melrose, D. B., Robinson, P. A., \& Feletto, T. M.
{1995}
%{Mode coupling due to twisting of magnetic field lines}
Solar Phys. {158}, 139

\bibitem[\protect\citeauthoryear{White, Thejappa \& Kundu}{1992}]{WTK92}
White, S. M.,  Thejappa, G.  \& Kundu, M. R. 1992
Solar Phys. {138}, 163


\bibitem[\protect\citeauthoryear{Zheleznyakov}{1970}]{Z70}
Zheleznyakov, V. V. 1970, {\it Radio emission from the Sun and planets}, Pergamon Press, New York

\bibitem[\protect\citeauthoryear{Zheleznyakov \& Zlotnik}{1964}]{ZZ64}
Zheleznyakov, V. V. \& Zlotnik, E. Ya. 1964, Sov. Astron. AJ 7, 485
\end{thebibliography}
\end{document}